\documentstyle[12pt]{article}

\newcommand{\be}{\begin{equation}}
\newcommand{\ee}{\end{equation}}
\newcommand{\ba}{\begin{eqnarray}}
\newcommand{\ea}{\end{eqnarray}}

\newcommand{\tr}{\rm tr}

\begin{document}
\setlength{\textheight}{8.5 in}
\begin{titlepage}
\begin{center}
\hfill {\bf{ LPTENS 12/46}}\\
\hfill {\bf Dec.20, 2012 }\\
\vskip 0.6 in
{\large \bf{The intersection numbers of the $p$-spin curves from random matrix theory}}
\vskip .6 in
\begin{center}
 {\bf E. Br\'ezin$^{a)}$ and S. Hikami$^{b)}$}
\end{center}
\vskip 5mm
\begin{center}
{$^{a)}$ Laboratoire de Physique Th\'eorique, Ecole Normale Sup\'erieure}\\ {24 rue Lhomond 75231, Paris
Cedex
05, France. e-mail: brezin@lpt.ens.fr{\footnote{\it
Unit\'e Mixte de Recherche 8549 du Centre National de la
Recherche Scientifique et de l'\'Ecole Normale Sup\'erieure.
} }}\\
{$^{b)}$ Mathematical and Theoretical Physics Unit, OIST Graduate University,\\
1919-1 Tancha, Onna-son, Okinawa 904-0495  Japan. e-mail: hikami@oist.jp}\\
\end{center}     
\vskip 3mm         

{\bf Abstract}                  
\end{center}

The intersection numbers of  $p$-spin curves are computed through  correlation functions of 
Gaussian ensembles of random matrices in an external matrix source. The $p$-dependence of  intersection numbers 
is determined as  polynomial in $p$ ; the   large  $p$ behavior  is also considered. The analytic continuation of 
intersection numbers to  negative values of $p$ is  discussed  in   relation to SL(2,R)/U(1) black hole sigma
model.

\end{titlepage}
\vskip 3mm

\section{Introduction}

The intersection numbers for  $p$-spin curves appear in 
the generalized Kontsevich matrix model \cite{Kontsevich,Witten,Witten1}.  The generating function for $p$-spin intersection number obeys the $p$-th KdV equation or Gelfand-Dikii equation. In a random matrix theory, the  correlation functions at the edges of the spectrum, where one can tune a degeneracy of order $p$,
are expressed through intersection numbers \cite{Okounkov,BH1,BH2,BH3}.  In conformal field theory, the $p$-spin curve intersection theory  is related to $\mathcal{N}$=2 superconformal  minimal theory  for Lie algebra $A_{p-1}$ type.  It has been pointed out that  it corresponds to a gauged Wess-Zumino-Witten (WZW) model  of $SU(2)_{k}/U(1)$, where $k=p-2$ is the level of the Kac-Moody algebra of Lie group $SU(2)$ \cite{WZW}, and is related to SL(2,R)/U(1) black hole sigma model
when $k$ becomes negative \cite{WittenBlack}.

 The free energy for the $p$-spin curve satisfies interesting universal equations,  such as a string equation, dilaton equation, and WDVV equation, so called tautological equations or universal equations, and has been studied in the connection to Gromov-Witten theory \cite{Jarvis,Chiodo,Faber}.
Although the $p$-spin curve intersection numbers can be obtained through tautological equations  in a recursive way, the actual computation for higher genuses  is limitted  \cite{LX1,LX2,KL}. 

In  previous papers, we have derived explicit integral formula for the $p$-spin curve intersection numbers  of the moduli space
$\overline {M}_{g,n}$ valid for all order of genus $g$. We have shown that they are obtained analytically for a fixed number $n$ of marked  points.
Our  formulation starts from  simple Gaussian
matrix models with an external matrix source  and based upon a duality relation \cite{BH1,BH2,BH3}, from which one recovers a generalized Kontsevich matrix model.

The intersection numbers for the spin moduli spaces with 
$n$-marked points  are obtained
from the n  $n$-point correlation functions $U(s_1,...,s_n)$
of Gaussian random matrices in a scaling limit at critical edges \cite{BH8,BH9}. 
In a previous article \cite{BH10}, we  have computed explicitly 
the intersection numbers of moduli space of $p$-spin curves
with one marked point, for arbitrary values of $p$, as  polynomials in $p$.
This allowed us to consider continuations in $p$ ; in particular 
the limit $p\to -1$  exhibits an interesting relation between the intersection numbers,
 and the orbifold Euler characteristics $\chi(\overline{M}_{g,1})=
\zeta( 1- 2g)$, where $\zeta(x)$ is the Riemann zeta function) \cite{HZ,Penner}.

In this paper, we  extend the  evaluation of  the intersection numbers beyond the one marked point ($n=1$)
for arbitrary $p$. The obtained intersection numbers are consistent with previously known results \cite{LX1,LX2,KL} for small values of
$p$.
We pursue  the large $p$ behavior,    $p\to \infty$ limit.  The $p$-spin curve intersection theory 
is equivalent to  gauged WZW model. For this gauged WZW theory, in which $k=p-2$ appears as overall factor, the large $k$ limit may give  a semi-classical solution  \cite{WittenBlack,Kounnas}.   In the negative $k$, the gauged WZW model 
on $SU(2)/U(1)$ is changed to WZW model on non-compact $SL(2,R)/U(1)$, which is relevant to a black hole $\sigma$ model \cite{WittenBlack}. We will discuss the relation between the intersection numbers and  the density of state of $SL(2,R)/U(1)$ black hole sigma model
\cite{Hanany,Maldacena,Ikhlef}.

\section{Generating function for $p$-spin Intersection numbers }
The mathematical definition of the intersection numbers of the moduli space of $p$-spin curves with $s$-marked points is given by\cite{Witten1}
\be\label{Uj1}
< \tau_{n_1}(U_{j_1}) \cdots \tau_{n_s}(U_{j_s}) > = 
\frac{1}{p^g}\int_{\overline{M}_{g,s}} C_T(\nu) \prod_{i=1}^s
(c_1 ({\mathcal L}_i))^{n_i}
\ee
where $U_j$ is an operator for the primary matter field (tachyon), related to 
top Chern class $C_T(\nu)$, and 
$\tau_{n}$ is a gravitational operator,
related to the first Chern class $c_1$ of the line bundle ${\mathcal L}_i$ 
at the $i$th-marked point.
We denote $\tau_{n}(U_j)$ by $\tau_{n,j}$, and $j$ represents the spin index (j=0,...,p-1).
The problem of definition (\ref{Uj1}) has been discussed extensively \cite{Chiodo}.

In a previous paper \cite{BH10}, we have shown that those intersection numbers  (\ref{Uj1}) are expressed trhough the
correlation functions $U(s_1,...,s_n)$ as coefficients of powers of $s_j$,
\ba\label{U1m}
U(s_1,s_2,...,s_n) &=& < {\tr} e^{s_1 M} {\tr}e^{s_2 M} \cdots {\tr}e^{s_n M}>\nonumber\\
&=& \int \prod_{l=1}^m d \lambda_l e^{\sum i t_l \lambda_l} 
<\prod_1^m  {\tr } \delta (\lambda_j - M) >
\ea
where $s_l= i t_l$ ;  $M$ is an $N\times N$ Hermitian random matrix.
The bracket stands for averages with the Gaussian probability measure 
\be
 < X > =\frac{1}{Z} \int dM e^{-\frac{N}{2}{\tr} M^2 +N {\tr }M A} X(M),
\ee
in which $A$ is an $N\times N$ external Hermitian  matrix source. By an appropriate  tuning of the
external source matrix $A$, we may obtain the desired singularity, which generates the $p$-spin curves.
The relation to the generalized Kontsevich model is discussed  in $\S$ 3,4 of \cite{BH10}.

An exact and  useful integral representation for $U(s_1,...,s_n)$ is  known
in  presence of an arbitrary external 
matrix source $A$ with eigenvalues $a_{\alpha}$ \cite{BH5} : 
\ba\label{U(sn)} 
&&U(s_1,\cdots, s_n) = \frac{1}{N} \langle \rm{tr} e^{s_1 M} \cdots \rm{tr} e^{s_n M} 
\rangle  \nonumber\\
&&= e^{\sum_1^n s_i^2}\oint \prod_1^n \frac{du_i}{2\pi i} 
e^{\sum_1^n u_is_i} \prod_{\alpha=1}^N \prod_{i=1}^n (1-\frac{s_i}{a_\alpha- u_i}) 
\det \frac{1}{u_i-u_j + s_i}
\ea

This representation involves  contour integrals around $u_i= a_\alpha$.  In the large N limit, it is convenient to
express the factors in the determinant as additional integrals. For instance, in the case of the two point correlation (n=2), after the shift
$u_i\to u_i-\frac{s_i}{2}$, $s_i\to \frac{s_i}{N}$, in the two point function, we have
\ba
&&\frac{1}{u_1-u_2 + \frac{1}{2N}(s_1+s_2)} \frac{1}{u_1-u_2- \frac{1}{2N}(s_1+s_2)}\nonumber\\
&&= \frac{N}{s_1+s_2} \int_0^\infty dx e^{-x(u_1-u_2)}{\rm sh} (\frac{x}{2N}(s_1+s_2))
\ea
Tuning  now the $a_\alpha$'s, and taking the large N limit, we obtain
\ba\label{U2}
&&U(s_1,s_2) = \frac{2N}{s_1+s_2}\frac{1}{(2\pi i)^2}\int_0^\infty dx
\int du_1 du_2 {\rm sh}(\frac{1}{2N} x (s_1+s_2))e^{-(u_1-u_2)x}\nonumber\\
&&\times{\rm exp}[ - \frac{N}{p^2-1}\sum \frac{1}{a_\alpha^{p+1}} 
(\sum_i (u_i + \frac{1}{2N}s_i)^{p+1}
-\sum_i (u_i-\frac{1}{2N}s_i)^{p+1}) ]
\ea
For the three and four point correlations, similar useful formulae for the determinant part of (\ref{U(sn)}) may be found in the 
appendices A and B of \cite{BH10}. 
\section{ Intersection numbers for $p=3$  with two marked points}

The intersection numbers are obtained as coefficients of the power series in ${s_1,s_2}$ of  $U(s_1,s_2)$.
In a previous paper \cite{BH10}, for p=3,  we have computed the intersection numbers  with two marked points  or genus one case (g=1) starting from(\ref{U2}). As an example, we compute the $p=3$ case up to genus 3.
The general expansion
\be\label{U2}
U(s_1,s_2) =\sum_{g,m,j} < \tau_{m_1,j_1}\tau_{m_2,j_2}>_g \Gamma(1 - \frac{1+j_1}{p})\Gamma(1 - \frac{1+j_2}{p}) {s_1}^{m_1^\prime}
{s_2}^{m_2\prime}
\ee
with the condition,
\be
(p+1)(2g - 2 + n) = \sum_{i=1}^{s}(p m_i + j_i + 1),\hskip 2mm m_k^\prime =  m_k + \frac{1+j_k}{p}\hskip 2mm (k=1,2)
\ee
is applied to the special case $n=2,p=3$ The gamma functions in (\ref{U2}) represent the spin factors.

After rescaling of  the parameters, 
\ba\label{U12}
U(s_1,s_2) &=&\frac{2}{(s_1+ s_2)(3 s_2)^{1/3}} \int_0^\infty dy {\rm sh}(\frac{s_1+s_2}{2}(3 s_1)^{1/3}y)
A_i (y - \frac{1}{4\cdot 3^{1/3}}{s_1}^{8/3})\nonumber\\
 &&\times A_i(- a y - \frac{1}{4\cdot 3^{1/3}}{s_2}^{8/3})
\ea
in which  $a = (s_1/s_2)^{1/3}$, and the Airy function is 
\be
A_i(y) = \int_{-\infty}^{+\infty} \frac{du}{2\pi}  e^{\frac{i}{3}u^3 + i u y} 
\ee

The Airy function satisfies the differential equation 
\be\label{Airy}
A_i^{\prime\prime}(y) =  y A_i(y), \hskip 3mm 
A_i^{\prime\prime}( -a y) = - a^3 y A_i(- a y)
\ee
The genus one case  (g=1) has been studied in \cite{BH3}.

If one expands  the hyperbolic sine function and the Airy functions in (\ref{U12})  up to relevant orders, we find a sum of six terms which, for $g=2$, 
involve the following integrals: 
\ba\label{integralI}
&&  I_1 = \int_0^\infty dy y^5 A_i(y)A_i(-a y),\hskip2mm I_2= \int_0^\infty dy  y A_i^{\prime\prime}(y) A_i(-a y),\nonumber\\
 && I_3= \int_0^\infty dy y A_i(y)A_i^ {\prime\prime}(- a y),\hskip 2mm
  I_4 = \int_0^\infty dy y A_i^\prime(y)A_i^\prime(-a y),\nonumber\\
  &&I_5= \int_0^\infty dy  y^3 A_i^{\prime} (y) A_i(-a y),
  \hskip2mm I_6= \int_0^\infty dy y^3 A_i(y) A_i^ {\prime}(- a y)
  \ea
  A repeated use of (\ref{Airy}) plus integrations  by parts allows us to  write all these integrals in terms of 
 \be A_i(0) = \frac{ 3^{-2/3}}{\Gamma(2/3)} 　= \frac{1}{2\pi 3^{1/3}}\Gamma(\frac{1}{3}),\hskip1cm  A_i^{\prime}(0) = -\frac{ 3^{-1/3}}{\Gamma(1/3)}=-\frac{1}{2\pi}\Gamma(\frac{2}{3}) \ee
 plus the integral
   \be\label{T}
  T = \int_0^\infty dy A_i(y) A_i^\prime (- ay)
  \ee
  which cannot be reduced to $A_i(0)$ or $A_i^{\prime}(0)$. 
  For instance one finds 
  \be (1+a^3) I_2 = A_i(0)^2 -2T \ee
  and so on. 
  However, all the T-dependence cancels when we sum up
  all the terms relevant to $g=2$ in $U(s_1,s_2)$. For instance the sum of all terms of order  $s_2^{\frac{16}{3}} $
is given by 

\ba
 &&\frac{1}{5!}\frac{1}{16} 3^{4/3} (1 + a^3)^4 a^5 s_2^{\frac{16}{3}} I_1
+\frac{1}{2}(\frac{1}{4\cdot 3^{1/3}})^2  a^{17}s_2^{\frac{16}{3}} I_2\nonumber\\
&&+\frac{1}{2}(\frac{1}{4\cdot 3^{1/3}})^2a^{-1} s_2^{\frac{16}{3}} I_3
 - (\frac{1}{4\cdot 3^{1/3}})^2 a^8 s_2^{\frac{16}{3}} I_4\nonumber\\
&& -\frac{1}{3!}\frac{1}{16}3^{1/3}(1+a^3)^2 a^{11} s_2^{\frac{16}{3}} I_5
+\frac{1}{3!}\frac{1}{16} 3^{1/3}a^2 (1+ a^3)^2 s_2^{\frac{16}{3}}  I_6
\ea
 and we add up the six terms and expand in powers of $s_1$ to the relevant orders we find 
  
\ba\label{U100}
U(s_1,s_2)\vert_{g=2} \hskip 1mm=&& \frac{(A_i(0))^2}{32\cdot 3^{2/3}}( - s_1^{14/3}s_2^{2/3} - \frac{11}{5}s_1^{11/3}s_2^{5/3}
\nonumber\\
&& - \frac{17}{5}s_1^{8/3}s_2^{8/3} - \frac{11}{5}s_1^{5/3}s_2^{11/3} - s_1^{2/3}s_2^{14/3}).
\ea
From these  results,  we obtain the intersection numbers 

\ba\label{g2p3}
&& <\tau_{0,1}\tau_{4,1}>_{g=2} = \frac{1}{864}\nonumber\\
&&<\tau_{1,1}\tau_{3,1}>_{g=2} = \frac{11}{4320}\nonumber\\
&&<\tau_{2,1}\tau_{2,1}>_{g=2} = \frac{17}{4320}
\ea

Rather than computing the exact dependence in $a$ of the terms proportional to $s_2^{16/3}$ and then re-expand in $a $ to obtain the various terms of (\ref{U100}),  we may proceed in a simpler way by expanding $A_i(-ay),A_i^\prime(-ay),A_i^{\prime\prime}(-ay)$ for small $a$:
\be
A_i(-ay) = A_i(0) - a y A_i^\prime (0) + \frac{a^2}{2}y^2 A_i^{\prime\prime}(0) +
\cdots
\ee 
and we then recover  (\ref{g2p3}).

In the genus- three case (g=3), we have again ten distinct integrals  $J_1 -  J_{10}$ for the terms of order $s_2^8 a^m$, in the small $s_1,s_2$ expansion of (\ref{U12}).
\ba
&&J_1=\int_0^\infty dy y^7 A_i(y)A_i(- a y) ,\hskip 2mm J_2=\int_0^\infty dy y^5 A_i^\prime (y)A_i(- a y)\nonumber\\
&&J_3=\int_0^\infty dy y^5 A_i(y)A_i^\prime(- a y),\hskip 2mm J_4=\int_0^\infty dy y^3 A_i^\prime (y)A_i^\prime(- a y)\nonumber\\
&&J_5=\int_0^\infty dy y^3 A_i^{\prime\prime}(y)A_i(- a y),\hskip 2mm J_6=\int_0^\infty dy y^3 A_i (y)A_i^{\prime\prime}(- a y)\nonumber\\
&&J_7=\int_0^\infty dy y A_i^{\prime\prime\prime}(y)A_i(- a y),\hskip 2mm J_8=\int_0^\infty dy y A_i (y)A_i^{\prime\prime\prime}(- a y)\nonumber\\
&&J_9=\int_0^\infty dy y A_i^{\prime\prime}(y)A_i^\prime(- a y),\hskip 2mm J_{10}=\int_0^\infty dy y A_i ^\prime(y)A_i^{\prime\prime}(- a y)
\ea
The genus 3 contribution for $U(s_1,s_2)$ is then expressed as the sum of four terms, $U^{(1)}$ -  $U^{(4)}$.
The term $U^{(1)}$, which is related to $J_1$, is  
\ba\label{U1}
 U^{(1)} &= &\frac{9}{7!\cdot 64}s_1^{7/3}s_2^{17/3} (1+a^3)^6 J_1\nonumber\\
 &= &\frac{3}{8960}s_2^8(-15 a^7 -42 a^8 + 90 a^{10}+ 63 a^{11}\nonumber\\
&& + 63 a^{13} + 90 a^{14} -42 a^{16} - 15 a^{17}) A_i(0)A_i^\prime(0).
\ea
The term $U^{(2)}$, which is related to  $J_2$, is
\ba
U^{(2)} &= &- \frac{1}{2560}s_2^8 a^{13}(1 + a^3)^4 J_2\nonumber\\
&=& -\frac{1}{2560}s_2^8 a^{13}(30 + 72 a -120 a^3 - 90 a^4 + 12 a^6)A_i(0)A_i^\prime(0).
\ea
The term $U^{(3)}$, which is related to $J_3$, is
\ba
U^{(3)} &=& \frac{1}{2560} s_2^8 a^4 (1+ a^3)^4 J_3\nonumber\\
&=&\frac{1}{2560}s_2^8 a^5 (-12 + 90 a^2 + 120 a^3 - 72 a^5 - 30 a^6)A_i(0)A_i^\prime(0).
\ea

The term $U^{(4)}$ from the sum of the contributions of $J_4$ to $J_{10}$.
We have
\ba
J_{10} &=&\frac{1}{1+a^3}( 2 a^3 K_1 - a^3 K_2)\nonumber\\
J_9 &=& - K_2 - J_{10}\nonumber\\
J_8 &=&\frac{a^3 + 2 a^4 - 2 a^6 - a^7}{(1 + a^3)^2} L\nonumber\\
J_7 &=&\frac{1}{a^3} J_8\nonumber\\
J_6 &=&\frac{6 a^3}{1 + a^3} J_7\nonumber\\
J_5 &=&-\frac{1}{a^3} J_6\nonumber\\
J_4 &=& a^3 J_5 - 3 J_9
\ea
with $L= A_i(0)A_i^\prime (0)$, and $K_1$,$K_2$ are given below.
We have also the  following relations between $J_1,J_2$ and $J_3$,
\ba
J_1 &=&\frac{1}{1+a^3}(30 J_5 + 12 J_3)\nonumber\\
J_2 &=&\frac{1}{1+a^3}(-5 J_5 + 4 J_4)\nonumber\\
J_3 &=&\frac{1}{1+ a^3}(- 5 a^3 J_5 - 4 J_4)
\ea
Thus $U^{(4)}$ becomes
\ba
U^{(4)}& =& \frac{1}{1152} s_2^8 ( a + 5 a^4 - 20 a^7 + 23 a^{10}+ 16 a^{13}\nonumber\\
&& -19 a^{16} + 13 a^{19} + 2 a^{22} + ( 2 a^2 + 13 a^5 - 19 a^8\nonumber\\
&&+ 16 a^{11} + 23 a^{14} - 20 a^{17} + 5 a^{20} + a^{23}))A_i(0)A_i^\prime(0).
\ea
Since $a= (s_1/s_2)^{1/3}$, the above expression for $U(s_1,s_2)$ is a symmetric function in $s_1$ and $s_2$.
Denoting 
\be
s^{m+\frac{1+j}{3}} = t_{m,j}
\ee
and
dividing $U(s_1,s_2)$ by $1/g^p$,  i.e. $1/27$ in this case, we obtain the intersection numbers 
$<\tau_{m_1,j_1}\tau_{m_2,j_2}>_{g=3}$ as the coefficient of $t_{m_1,j_1}t_{m_2,j_2}$ in $U(s_1,s_2)$ taking into account the spin factors.
The following spin factor appears as a over all factor in $U(s_1,s_2)$ at genus 3.
\be\label{Ai0}
A_i(0)A_i^\prime(0) = - \frac{1}{(2\pi)^2 3^{1/3}} \Gamma(\frac{1}{3})\Gamma(\frac{2}{3}) 
\ee
where $\Gamma(\frac{1}{3})$, $\Gamma(\frac{2}{3})$ are spin 1 and spin 2 factors, respectively, as (\ref{U2}).
All  the integrals $J_1,...,J_{10}$ are expressed by (\ref{Ai0}), and there are no terms like (\ref{T}), which appeared in
the integrals for the g=1,g=2 cases. Finally we have to compute the following terms 
\ba
K_1 &=&\int dy A_i^{\prime\prime}(y)A_i(-a y) = -A_i(0)A_i^\prime (0) - K_2\nonumber\\
K_2 &=&\int dy A_i^\prime (y) A_i^\prime (- a y).
\ea
For these integrals, we find
\be
K_1 = -\frac{1 + a}{1+a^3}A_i(0)A_i^\prime (0), \hskip 2mm K_2 =\frac{a - a^3}{1 + a^3} A_i(0)A_i^\prime (0)
\ee
and all the integrals reduces to the spin factor (\ref{Ai0}).
Summing up the results of $U^(1)$ to $U(4)$, we obtain the intersection numbers for $p=3, g=3$,
\ba
&&<\tau_{0,0}\tau_{7,1}>_{g=3} = \frac{1}{31104},\hskip2mm <\tau_{0,1}\tau_{7,0}>_{g=3} = \frac{1}{15552}\nonumber\\
&&<\tau_{1,0}\tau_{6,1}>_{g=3} = \frac{5}{31104},\hskip2mm <\tau_{1,1}\tau_{6,0}>_{g=3} = \frac{19}{77760}\nonumber\\
&&<\tau_{2,0}\tau_{5,1}>_{g=3} = \frac{103}{217728},\hskip2mm < \tau_{2,1}\tau_{5,0}>_{g=3} = \frac{47}{77760}\nonumber\\
&&<\tau_{3,0}\tau_{4,1}>_{g=3}=\frac{443}{544320},\hskip2mm < \tau_{3,1}\tau_{4,0}>_{g=3}=\frac{67}{77760}
\ea
The above results  are in complete agreement with the previous results \cite{LX1,KL}, which were obtained by  recursion relations.

\section{Intersection numbers for $p >3$}

For higher multicritical points  the algebra is similar, except that we have to deal with generalized Airy functions. For instance for $p=4$ instead of $A_i(x)$ we have to work with $\phi(x)$ defined as
\be
\phi(x) = \int_{0}^\infty dv e^{-\frac{1}{4} v^4 + v x}
\ee
which satifies
\be\label{phi4}
\phi^{\prime\prime\prime}(x) = x \phi .
\ee
Then, similarly
\ba
U(s_1,s_2) &=& \frac{2}{(s_1+s_2)(4 s_2)^{1/4}}\int_0^\infty dx \int_0^\infty dv_1 dv_2
{\rm sh}(\frac{s_1+s_2}{2}(4 s_1)^{1/4} x) \nonumber\\
&&e^{-\frac{s_1^3}{2}(\frac{1}{4 s_1})^{1/2} v_1^2
- \frac{s_2^3}{2}(\frac{1}{4 s_2})^{1/2} v_2^2}
e^{-\frac{1}{4}v_1^4 + x v_1 -\frac{1}{4} v_2^4 - a x v_2}
\ea
where $a = (s_1/s_2)^{1/4}$.
In complete analogy with the $p=3$ case, a  repeated use of  integration by parts and of  (\ref{phi4}) leads  to the expansion of $U(s_1,s_2)$.  In the genus one case,
\ba 
U(s_1,s_2)\vert_{g=1}  &= &\frac{1}{4}(\phi^{\prime\prime}(0))^2 s_1^{1/4}s_2^{1/4}(s_1^2 + s_1 s_2 + s_2^2)\nonumber\\
&&+ \frac{1}{12}(s_1s_2)^{3/4}(s_1+s_2)(\phi(0))^2
\ea
with
\be 
\phi^{\prime\prime}(0) = 2^{1/2} \Gamma(\frac{3}{4}), \hskip 2mm \phi(0) = 2^{-1/2}\Gamma(\frac{1}{4})
\ee
which provide the  $j=0,j=2$ spin factors respectively.
Replacing $s_1,s_2$ by $t_m,j$, ($s^{m+(1+j)/p} = t_{m,j}$),
\ba\label{p4g1}
U(s_1,s_2)\vert_{g=1} &=& \frac{1}{2}(t_{2,0}t_{0,0} + t_{1,0}t_{1,0} + t_{0,0}t_{2,0})(\Gamma(\frac{3}{4}))^2\nonumber\\
&&+ \frac{1}{24}(t_{1,2}t_{0,2} + t_{0,2}t_{1,2})(\Gamma(\frac{1}{4}))^2
\ea
Multiplying by  a factor $\frac{1}{p^g}$, we obtain the intersection numbers as coefficients of (\ref{p4g1}) for $p=4$ in the genus
one case,
\be\label{p4}
<\tau_{0,0}\tau_{2,0}>_{g=1} = \frac{1}{8},\hskip 2mm <\tau_{1,0}\tau_{1,0}>_{g=1}= \frac{1}{8},\hskip 2mm
<\tau_{0,2}\tau_{1,2}>_{g=1} = \frac{1}{96}
\ee
For $g=2$, $p=4$, from the term $s_2^\frac{18}{4}s_1^\frac{2}{4}$, we have similarly
\be\label{g2p4}
<\tau_{0,1}\tau_{4,1}>_{g=2} = \frac{1}{320}
\ee
 
For general $p$ we have to deal with the generalized Airy functions $\phi(x)$ for $p>2$, which satisfy the differential equation,
\be
\phi^{(p)} (x) = x \phi(x)
\ee
where $\phi^{(p)}(x)$ means the $p$-th derivative of $\phi(x)$. The generalized Airy function has an integral representation,
\be
\phi(y) = \int_0^\infty du\hskip 1mm e^{-\frac{u^{p}}{p} + y u}.
\ee
As examples of what the method can provide we give a few results : 
 for the case $p=5$, we obtain
 \ba\label{p5}
 &&<\tau_{1,3}\tau_{0,2}>_{g=1} = <\tau_{1,2}\tau_{0,3}>_{g=1} = \frac{1}{60}\nonumber\\
 &&<\tau_{1,0}\tau_{1,0}>_{g=1} = <\tau_{0,0}\tau_{2,0}>_{g=1} = \frac{1}{6}\nonumber\\
 &&<\tau_{0,1}\tau_{4,1}>_{g=2} = \frac{7}{1200}.
 \ea
 For the case $p=6$, 
 \be\label{p6}
 <\tau_{0,3}\tau_{1,3}>_{g=1} = \frac{1}{36},\hskip 2mm <\tau_{0,2}\tau_{1,4}>_{g=1}=\frac{1}{48},
 \hskip 2mm <\tau_{0,4}\tau_{1,2}>_{g=1}=\frac{1}{48}.
 \ee
 
 For the case $p=7$, 
 \ba\label{p7}
 &&<\tau_{0,2}\tau_{1,5}>_{g=1} = <\tau_{1,2}\tau_{0,5}>_{g=1} =\frac{1}{42}\nonumber\\
 &&<\tau_{0,4}\tau_{1,3}>_{g=1} = <\tau_{0,3}\tau_{1,4}>_{g=1} =\frac{1}{28}\nonumber\\
 &&<\tau_{1,0}\tau_{1,0}>_{g=1} = <\tau_{0,0}\tau_{2,0}>_{g=1} = \frac{1}{4}.
 \ea

\section{The $p$ dependence of the intersection numbers}

In a previous article \cite{BH10}, we have considered the intersection numbers with  one marked point  
for arbitrary $p$, and found results such as 
\ba\label{pdepen}
<\tau_{1,0}>_{g=1} =&& \frac{p-1}{24}\nonumber\\
<\tau_{n,j}>_{g=2} =&& \frac{(p-1)(p-3)(1 + 2 p)}{p\cdot 5!\cdot4^2\cdot 3}\frac{\Gamma(1-\frac{3}{p})}{
\Gamma(1-\frac{1+j}{p})}\nonumber\\
<\tau_{n,j}>_{g=3} =&& \frac{(p-5)(p-1)(1+2p)(8p^2-13 p -13)}{p^2\cdot 7!\cdot 4^3\cdot 3^2}
\frac{\Gamma(1-\frac{5}{p})}{\Gamma(1 - \frac{1 + j}{p})}\nonumber\\
<\tau_{n,j}>_{g=4} =&& \frac{(p-7)(p-1)(1+2 p)(72p^4-298p^3-17p^2+562p+281)}{p^3\cdot 9!\cdot 4^4\cdot 15}\nonumber\\
&&\times \frac{\Gamma(1-\frac{7}{p})}{\Gamma(1 -\frac{1+j}{p})}
\ea
with $n= 2 g -1 +\frac{2g-2-j}{p}$.
In the large $p$ limit,  the intersection numbers $<\tau_{n,j}>_g$ behave as
\be\label{largep}
<\tau_{n,j}>_g = \frac{B_g}{(2g)!(2g)} p^g + O(p^{g-1})
\ee
with $B_g$ is a  Bernouilli number, $B_1=\frac{1}{6},B_2=\frac{1}{30}, B_3=\frac{1}{42},B_4=\frac{1}{30}$.
Note the well known relation to $\zeta(2g)$ as
\be
\frac{B_g}{(2g)! (2g)} = \frac{1}{(2\pi )^{2g} g} \zeta(2g)
\ee

We have derived (\ref{largep}) from $U(s)$ in the large $p$ limit. The one point function $U(s)$ has the following
expression\cite{BH8},
\be\label{U(s)}
U(s) =\frac{1}{Ns}\int\frac{du}{2i\pi}{\rm exp}(-\frac{c}{p+1}((u +\frac{1}{2}s)^{p+1}-(u-\frac{1}{2}s)^{p+1})
\ee
With $s = \frac{\sigma}{p}$, and $u^{p+1}=x^2$, we have
\be
U(s) = \frac{2}{N\sigma}\int \frac{dx}{2i\pi} x^{-1 +\frac{2}{p}}e^{-\frac{c}{p+1}x^2 (e^{\sigma/2}-e^{-\sigma/2})}
\ee
Thus we obtain
\be
U(s) = \frac{2}{N\sigma}\Gamma(\frac{2}{p}) (\frac{2c}{p+1}{\rm sh}\frac{\sigma}{2})^{-1/p}
\ee
This may be written as
\be
U(s) = \frac{2}{N\sigma}\Gamma(\frac{2}{p}) (\frac{2c}{p+1})^{-\frac{1}{2}}(\frac{\sigma}{2})^{-\frac{1}{p}}
{\rm exp}(-\frac{1}{p} {\rm log}\frac{ {\rm sh}\frac{\sigma}{2}}{\frac{\sigma}{2}})
\ee
and
expanding the exponent in $\frac{1}{p}$,  we find
\be
U(s) = \frac{2}{N\sigma} \Gamma(\frac{2}{p})(\frac{2c}{p+1})^{-\frac{1}{2}}(\frac{\sigma}{2})^{-\frac{1}{p}} (  1 - \frac{1}{p}{\rm log }(\frac{{\rm sh}(\frac{\sigma}{2})}{(\frac{\sigma}{2})}))
\ee
If we  use the expansion
\be\label{formula}
{\rm log}(\frac{{\rm sh} \frac{\sigma}{2}}{\frac{\sigma}{2}}) = \sum_{n=1}^\infty (-1)^{n-1}\frac{B_n  \sigma^{2n}}{(2n)! 2n}
\ee
and drop the  factors $(\frac{2c}{p+1})^{-\frac{1}{p}}$, and $(\sigma/2)^{-1/p}$ which are close to one in the large $p$ limit, we obtain 

\be\label{Bernoulli}
U(s) = (1 - \frac{1}{p}\sum_{n=1}^\infty (-1)^{n-1}\frac{B_n}{(2n)! 2n} \sigma^{2n})\frac{ p}{N\sigma}\Gamma(1+\frac{2}{p})
\ee

Since the intersection numbers $<\tau_{n,j}>_g$ are related to $U(s)$ by \cite{BH8}
\be
U(s) = \sum_g <\tau_{n,j}>_g \frac{1}{N\pi}\Gamma(1-\frac{1+j}{p})s^{(2g-1)(1+\frac{1}{p})}p^{g-1} 
\ee
with $(p+1)(2g-1) = pn+j+1$,
we have rederived  the large $p$ behavior of (\ref{largep}). 

From  (\ref{formula}),  taking a derivative  with respect to  $\sigma$, gives,
\be
 \frac{1}{e^\sigma -1} +\frac{1}{2} - \frac{1}{\sigma} = \sum_{n=1}^\infty (-1)^{n-1} \frac{B_n}{(2n)!} \sigma^{2n-1},
 \ee
 
 Using this relation one obtains
\be\label{formula1}
\frac{d}{d\sigma} ( \sigma U(\sigma) )= \frac{1}{\sigma} -\frac{1}{2} -\frac{1}{e^{ \sigma} -1}
\ee

The di-gamma function $\psi(z)$ has the following expression,
\be\label{formula2}
\psi(z)=\frac{d}{d z} {\rm log}\Gamma(z) = {\rm log} z -\frac{1}{2z} -\int_0^\infty d \sigma (\frac{1}{2}-\frac{1}{\sigma}+\frac{1}{e^\sigma-1}) e^{- \sigma z}.
\ee

From (\ref{formula1}) and (\ref{formula2}) we find thus in the large $p$ limit,
\ba\label{density}
\frac{d}{d z} {\rm log}\Gamma(z) = &&{\rm log} z -\frac{1}{2z}+  \int _0^\infty d\sigma  (\frac{d}{d \sigma} {(\sigma U(\sigma))} )e^{-  \sigma z}\\ \nonumber
 = && \hskip 1mm {\rm log }z -  \frac{1}{2 z} - z\frac{d}{dz} \int d\sigma U(\sigma)e^{-\sigma z}
 \ea
 The last integral  is related to the density of states. 
 In (\ref{U1m}), $s$ is replaced by $s=i t$, and if we replace $z$ by $i E$, and take the imaginary part, we obtain the density of states $\rho(E)$.
 After integration by parts, we obtain
 \be\label{density3}
 \rho(E) = \frac{d}{dE} {\rm Im}\hskip 1mm{\rm  log}\Gamma(i E) - \frac{\pi}{2} - \frac{1}{2  E}
 \ee
 
We will  discuss this expression   in  connection to  the  density  of  states of  the $SL(2,R)/U(1)$ black hole sigma model in the next section.\\

Next we consider the two point correlation function $U(s_1,s_2)$.
For general $p$, $U(s_1,s_2)$ is expressed  as
\ba\label{generalp}
&&U(s_1,s_2) = \frac{2}{(s_1+s_2)(ps_2)^{1/p}}\int_0^\infty dx
\int_0^\infty dv_1 dv_2 {\rm sh}(\frac{s_1+s_2}{2}(p s_1)^{1/p})\nonumber\\
&&e^{-\frac{v_1^p}{p} + x v_1 -\frac{p(p-1)}{24}s_1^3 (ps_1)^{\frac{2-p}{p}}v_1^{p-2} + \dots}
e^{-\frac{v_2^p}{p} - a x v_2 -\frac{p(p-1)}{24}s_2^3 (ps_2)^{\frac{2-p}{p}}v_2^{p-2} + \dots}
\ea
The exponent of (\ref{generalp}) follows from the binomial expansion,
\be
(u + \frac{s}{2})^{p+1} = u^{p+1} + (p+1)u^p (\frac{s}{2}) + (p+1)p \frac{1}{2}u^{p-1}(\frac{s}{2})^2 + \dots
\ee
and we use $c(p+1)=1$, $u^p s = v^p/p$. As in the case of p=3, polynomials  in $a$  (\ref{U1})
give the intersection numbers. Therefore we expand (\ref{generalp}) in power series of $a$.
At lowest order in  $a$, we obtain two  terms from (\ref{generalp}),
\ba
U_1 = &&\frac{1}{3!4}(s_1+s_2)^2\frac{(ps_1)^{\frac{3}{p}}}{(ps_2)^{\frac{1}{p}}}\int dx x^3 \phi(x)\phi(-a x)\nonumber\\
U_2= &&-\frac{p(p-1)}{24}(\frac{s_1}{s_2})^{\frac{1}{p}}s_2^3 (p s_2)^{\frac{2-p}{p}}\int dx x \phi(x)\phi^{(p-2)}(-a x)
(-a)^{2-p}
\ea
From $U_2$ we find a term  proportional to $a s_2^{2 +\frac{2}{p}}$ , namely
\be
\Delta U_2 = \frac{p-1}{24}p^{\frac{2}{p}} a s_2^{2 + \frac{2}{p}}(\phi^{(p-2)}(0))^2
\ee
with 
\ba
\phi^{(p-2)}(0) =&& \int_0^\infty du u^{p-2}e^{-\frac{u^p}{p}}\nonumber\\
=&&p^{-\frac{1}{p}}\Gamma(1 - \frac{1}{p}).
\ea

Since $s_2^{2 + \frac{1}{p}} s_1^{\frac{1}{p}} = t_{2,0}t_{0,0}$, we obtain
\be\label{p-1}
<\tau_{0,0}\tau_{2,0}>_{g=1} =\frac{p-1}{24}
\ee

From $U_1$ and $U_2$, we collect terms  proportional to $a^3 s_2^{2 + \frac{2}{p}}$ and obtain
\be\label{p-2}
<\tau_{0,2}\tau_{1,p-2}>_{g=1} = \frac{p-3}{24 p}
\ee
This result agrees with those obtained previously for $p=4,5,6$ and 7 in 
(\ref{p4}), (\ref{p5}), (\ref{p6})　and (\ref{p7}).
The intersection number (\ref{p-2}) can be neglected in the large $p$ limit in comparison with  (\ref{p-1}) .

Similarly  one obtains the g=2 terms from the coefficients of $a^m s_2^{4 + \frac{4}{p}}$ (m=1,2,3),
\ba\label{g2p}
&&< \tau_{0,0}\tau_{4,2}>_{g=2} =\frac{(p-1)(p-3)(2p + 1)}{5760 p}\nonumber\\
&&< \tau_{0,1}\tau_{4,1}>_{g=2} = \frac{(p-1)(p-2)(p + 2)}{2880 p}\nonumber\\
&&< \tau_{0,2}\tau_{4,0}>_{g=2} = \frac{(p-1)(p-3)(2p + 11)}{5760 p}
\ea
For the particular values of  $p=3,4,5$, the above expressions agree with the previous results (\ref{g2p3}),(\ref{g2p4}) and (\ref{p5}) for the genus two case.

From the  $a^5 s_2^{4+\frac{4}{p}}$ term, one finds
\be\label{gg2p}
<\tau_{0,4}\tau_{3,p-2}>_{g=2} = \frac{2 p^3 + 13 p^2 -158 p + 215}{5760 p^2}
\ee
which is valid for $p \ge 6$.

In the large $p$ limit,  the  three terms of (\ref{g2p}) become  equal, and coincide with the result for the one point intersection number (\ref{largep}).
\be
 <\tau_{0,m}\tau_{4,2-m}>_{g=2} = \frac{B_2 p^2}{4! \cdot 4}\hskip 2mm (p\to \infty)
 \ee
 
Note that (\ref{gg2p}) is order  $p$, and is negligible compared to (\ref{g2p}).

From the terms $a^m s_2^{6 + \frac{6}{p}}$ in the small $a$ expansion of $U(s_1,s_2)$, we obtain the g=3  (genus 3) terms.
In the case m=1, we have
\be
<\tau_{0,0}\tau_{6,4}>_{g=3} = \frac{(p-1)(p-5)(2p+1)(8p^2 - 13 p - 13)}{p^2\cdot 7! 4^3 3^2} \hskip 2mm(p > 5)
\ee
This is identical to $<\tau_{5,4}>_{g=3}$ in (\ref{pdepen}). The identity follows from the  string equation, in which
the insertion of  $\tau_{0,0}$ reduces the intersection number from  $s$ to $s-1$ marked points:   
\be
<\tau_{0,0}\prod_{i=1}^s \tau_{n_i,j_i}>_g = \sum_{l=1}^s <\tau_{n_l-1,j_l}\prod_{i=1,i\ne l}^s \tau_{n_i,j_i}>_g
\ee

In our formulation, this string equation follows from  the integral representation for the intersection numbers, 
when one collects  the terms proportional to $a$. By explicit calculation of two marked points, we verified this 
string equation. It might be possible to verify this string equation for n-marked points by the taking account of the term
of $a$.

From $a^2 s_2^{6 + \frac{6}{p}}$, we have for $p > 5$,
\be
<\tau_{0,1}\tau_{6,3}>_{g=3} = \frac{(p-1)(p-2)(p-4)(p+2)(2p+1)}{p^2 \cdot7! \cdot 8 \cdot 3^2}
\ee
From $a^3 s_2^{6 + \frac{6}{p}}$,
\be
<\tau_{0,2}\tau_{6,2}>_{g=3} = \frac{(p-1)(p-3)(16 p^3 + 34 p^2 - 155 p -129)}{p^2 \cdot7!\cdot 64\cdot 3^2}
\ee

In the large $p$ limit, these g=3 terms exhibit  same behavior as in (\ref{largep}),
\be
<\tau_{0,m}\tau_{6,4-m}>_{g=3} = \frac{B_3 }{6!\cdot 6} p^3 + O(p^2) \hskip 2mm (p\to\infty)
\ee

\section{ Analytic continuation to negative $p$}
One may analytically continue the integral representations of the correlation functions  to negative values of $p$. This continuation
was already examined in \cite{BH10}, and we recall some of the results here : 
\be\label{duallog}
U(s) = \frac{1}{Ns}\int \frac{du}{2i\pi} e^{-c
[
(u+\frac{1}{2}s)^{p+1} - (u - \frac{1}{2}s)^{p+1}]}
\ee
where $c= \frac{N}{p^2-1}\sum \frac{1}{a_\alpha^{p+1}}$.

Expanding the exponent, we obtain
\be\label{Bexpansion}
U(s) =\int \frac{du}{2 i \pi}{\rm exp}
[-c(s u^p + \frac{p(p-1)}{3! 4} s^3 u^{p-2}
+ \frac{p(p-1)(p-2)(p-3)}{5! 4^2}s^5 u^{p-4}+\cdots)].
\ee

This integrals yield Gamma functions 
after the replacement 
$u= (\frac{t}{c s})^{1/p}$,
\ba\label{Mexpansion}
U(s) &=& \frac{1}{Nsp}\cdot
\frac{1}{(c s)^{1/p}}\int_0^\infty dt
t^{\frac{1}{p}-1}e^{-(t + \frac{p(p-1)}{3! 4} s^{2+\frac{2}{p}}
c^{\frac{1}{p}} t^{1 - \frac{2}{p}}+
\frac{p(p-1)(p-2)(p-3)}{5! 4^2} s^{4+\frac{4}{p}} c^{\frac{4}{p}}
t^{1 - \frac{4}{p}}+\cdots)}\nonumber\\
&=&\frac{1}{Nsp}\cdot
\frac{1}{(c s)^{1/p}}
[ - \frac{p-1}{24} c^{\frac{2}{p}} y \Gamma(
1 - \frac{1}{p})+
\frac{(p-1)(p-3)(1+ 2 p)}{5! \cdot4^2\cdot3} 
y^2 \Gamma(1- \frac{3}{p})\nonumber\\
&-& \frac{(p-5)(p-1)(1+ 2 p)(8 p^2 - 13 p - 13)}{7! 4^3 3^2}
y^3 \Gamma(1 - \frac{5}{p}) \nonumber\\
&+& (p-7)(p-1)(1+ 2 p) (72 p^4 - 298 p^3 - 17 p^2 + 562 p + 281)
\nonumber\\
&\times& \frac{1}{9! 4^4 15} y^4 \Gamma(1 - \frac{7}{p}) \cdots]
\ea
with
$y = c^{\frac{2}{p}}s^{2 + \frac{2}{p}}$.
 
 From this expansion, we obtain the intersection numbers for one marked point as (\ref{pdepen}). 
 The intersection number $<\tau_{n,j}>_g$ is obtained from the term  $y^g \Gamma(1 - \frac{1}{p} - \frac{j}{
p})$ in (\ref{Mexpansion}).

The continuation  to $p<0$ is straightforward. The  $t$-integral  in (\ref{Mexpansion}) can be changed
to $v$ by $t = \frac{1}{v}$ ,$(0 < v < \infty)$, and one obtains the small $s$ expansion for negative $p$.
Therefore the expression for the intersection numbers (\ref{pdepen}) can be analytically continued to  negative $p$. This analytic continuation can also be done for two marked points, since we have
computed them in the previous sections for  general $p$.
For instance, from (\ref{pdepen}), we have the intersection numbers for $p= -3$,
\ba
&&<\tau_{1,0}>_{g=1} = -\frac{1}{6}, \hskip 2mm <\tau_{3,2}>_{g=2} = \frac{1}{144}\nonumber\\
&&<\tau_{6,1}>_{g=3} = -\frac{35}{34992}
\ea

In a  previous article \cite{BH10}, we have computed the intersection numbers $<\tau_{1,0}>_g$ for the case of $p=-1$ from $U(s)$, which provides the orbifold Euler characteristics $\chi({\mathcal M}_{g,1})$ with one marked point, 
\be\label{Euler}
<\tau_{1,0}>_g = \chi({\mathcal M}_{g,1}) = \zeta(1-2g) = - \frac{B_g}{2g}
\ee
with the Bernoulli number $B_g$, ($B_1= \frac{1}{6},B_2=\frac{1}{30},B_3=\frac{1}{42},...)$.
The s-point orbifold Euler characteristics $\chi({\mathcal M}_{g,s})$ may be obtained from the dilaton equation: 
\be
<\tau_{1,0} \tau_{n_1,j_1} \cdots \tau_{n_k,j_k}>_g = (2 g - 2 + k ) <\tau_{n_1,j_1}\cdots \tau_{n_k,j_k}>_g
\ee
Since the Euler characteristics with s marked points is $<\tau_{1,0}\cdots \tau_{1,0}>_g$,  the dilaton equation yields from (\ref{Euler}),
\be
\chi({\mathcal M}_{g,s}) = < (\tau_{1,0})^s >_g = - \frac{2g-1}{(2g)!}(2g + s -3)! B_g
\ee
This agrees with  previous results obtained in \cite{HZ,Penner,Bini}.

For $p=-2$, we have considered previously the equivalence with the unitary matrix model in a matrix source \cite{BH11}.

The central charge of the gauged Wess-Zumino-Witten model with symmetry $SU(2)_k/U(1)$ is 
\be
C = 2 - \frac{6}{k+2}
\ee
Changing $p$ to  $p= -p^\prime$,  $k$ to $k=-k^\prime$ ($p < 0$,$k < 0$), we have
$p^\prime = k^\prime - 2$, and the central charge $C$ is given by
\be
C = 2 + \frac{6}{k^\prime - 2}
\ee
The analytic continuation to negative
$p$ yields a gauged WZW model for  $SL(2,R)_{k^\prime}/U(1)$ . It is known that this model represents a  black hole $\sigma$ model\cite{WittenBlack}, in particular for the value $k^\prime =\frac{9}{4}\hskip1mm (p=-\frac{1}{4})$, for which the central charge $C$ becomes 26.

The density of states for the  $SL(2,R)/U(1)$ black hole has been studied in \cite{Hanany,Maldacena,Ikhlef},
\be\label{blackdensity}
\rho(E) = \frac{1}{\pi} {\rm log} \epsilon + \frac{1}{4\pi i}\frac{d}{d E} {\rm log}\frac{\Gamma(- iE +\frac{1}{2}- m)
\Gamma(- i E +\frac{1}{2} + \tilde m)}{\Gamma(+ iE +\frac{1}{2} + \tilde m)
\Gamma(+ i E +\frac{1}{2} -  m)}
\ee
in which $\epsilon$ is a regularization factor, and $m=\frac{1}{2}(n-k w)$, $\tilde m=-\frac{1}{2}(k w + n)$ are eigenvalues of $J_0^3$ and $\bar J_0^3$ in CFT $(J_0^3 - \bar J_o^3 = n,J_0^3 + \bar J_0^3 = - k w)$.
If we neglect $m$ , $\tilde m$, and the  $\frac{1}{2}$ terms in the large $E$ limit,
we obtain
\be
\rho(E) = \frac{1}{\pi} {\rm log} \epsilon + \frac{1}{2\pi i}\frac{d}{d E} {\rm log}\frac{\Gamma( -  iE )}{\Gamma( + iE )}
\ee
or
\be
 \rho(E) = \frac{2}{\pi}\frac{d}{d E} {\rm Im} \hskip 1mm {\rm log}\hskip 1mm\Gamma( -  i E)
 \ee

This expression agrees  with   (\ref{density3}), obtained from the intersection numbers for  large $p$.
We have  scaled $s = \sigma/p$, and the expression (\ref{density3}) is valid for small $s$. Therefore, the Fourier transform of $U(s)$ gives the large $E$ behavior, in which  the terms  $m$,$\tilde m$ and $1/2$ in (\ref{blackdensity}) can be
neglected.

\vskip 20mm


\section{ Discussion }
\vskip 5mm

    In this article, we have shown that the correlation functions $U(s_1,s_2,\cdots,s_n)$ of a Gaussian matrix model in a tuned external source, provide the intersection numbers
    for $p$-spin curves.  For instance, from the two point function $U(s_1,s_2)$, in the case of $p$=3, the intersection numbers
    are computed up to genus 3,  
    
    We  have also computed the intersection numbers for general   $p$. They are given by  power series in $a$,   $a= (\frac{s_1}{s_2})^{\frac{1}{p}}$. Then we have  considered the large $p$ behavior for  the two point functions.
    The density of states $\rho(E)$ becomes a di-gamma function in the large $p$ limit, and  this expression agrees with the density of states of a $SL(2,R)_k/U(1)$ WZW model, which has been studied in the context of two dimensional black hole solutions.
     The n-point correlation functions $U(s_1,\cdots,s_n)$ are known through the determinant of a kernel for the $p$-spin curve case. 
      It will be interesting to investigate further  the detailed comparison of those correlation functions, between $SL(2,R)_k/U(1)$ WZW theory and the intersection numbers for negative $p$-spin curves .
     
    \vskip 15mm

{\bf Acknowledgement}
\vskip 5mm
We thank Costas Kounnas for a comment on  negative $k$  for the $SU(2)_k/U(1)$ WZW model.

\end{document}